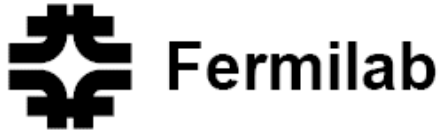

# SHIELDING EXPERIMENTS UNDER JASMIN COLLABORATION AT FERMILAB (III): MEASUREMENT OF HIGH-ENERGY NEUTROMS PENETRATING A THICK IRON SHIELD FROM THE ANTIPROTON PRODUCTION TARGET BY AU ACTIVATION METHOD[*][†]

H. Matsumura[1,#], N. Kinoshita[1,5], Y. Kasugai[2], N. Matsuda[2], H. Yashima[3], H. Iwase[1], A. Toyoda[1], N. Mokhov[4], A. Leveling[4], D. Boehlein[4], K. Vaziri[4], G. Lautenschlager[4], W. Schmitt[4], K. Oishi[5], Y. Sakamoto[2], and H. Nakashima[2]

[1]High Energy Accelerator Research Organization (KEK), 1-1 Oho, Tsukuba, Ibaraki 305-0801, Japan
[2]Japan Atomic Energy Agency, 2-4 Shirakata Shirane, Tokai-mura, Ibaraki 319-1195, Japan
[3]Kyoto University Research Reactor Institute, Kumatori-cho, Osaka 590-0494, Japan
[4]Fermi National Accelerator Laboratory, Batavia, IL 60510, USA
[5]Shimizu Corporation, 4-17, Etchujima 3-chome, Koto-ku, Tokyo 135-8530, Japan

## Abstract

In an antiproton production (Pbar) target station of the Fermi National Accelerator Laboratory (FNAL), the secondary particles produced by bombarding a target with 120-GeV protons are shielded by a thick iron shield. In order to obtain experimental data on high-energy neutron transport at more than 100-GeV-proton accelerator facilities, we indirectly measured more than 100-MeV neutrons at the outside of the iron shield at an angle of 50° in the Pbar target station. The measurement was performed by using the Au activation method coupled with a low-background γ-ray counting system. As an indicator for the neutron flux, we determined the production rates of 8 spallation nuclides ($^{196}$Au, $^{188}$Pt, $^{189}$Ir, $^{185}$Os, $^{175}$Hf, $^{173}$Lu, $^{171}$Lu, and $^{169}$Yb) in the Au activation detector. The measured production rates were compared with the theoretical production rates calculated using PHITS. We proved that the Au activation method can serve as a powerful tool for indirect measurements of more than100-MeV neutrons that play a vital role in neutron transport. These results will be important for clarifying the problems in theoretical calculations of high-energy neutron transport.

[*]Work supported by a Grant-in-Aid from the Ministry of Education (KAKENHI 19360432 and 20354764), Japan and by Fermi Research Alliance, LLC under contract No. DE-AC02-07CH11359 with the U.S. Department of Energy.
[†]Published in *Journal of the Korean Physical Society*.
[#]Corresponding author. E-mail : hiroshi.matsumura@kek.jp

## 1. INTRODUCTION

Various theoretical calculation codes are being used for designing the shields used in high-energy accelerator facilities. In 2006, computer simulations of the attenuation of the dose rate induced by neutrons incident at energies ranging from 20 MeV to 100 GeV on shield materials such as concrete and iron were performed using eight theoretical calculation codes. Then, the obtained results were compared and discussed at SATIF-8 (Shielding Aspects of Accelerators, Targets and Irradiation Facilities, Eighth Meeting, Korea) [1]. The difference among the obtained results was significant from the viewpoint of radiation protection. In particular, the difference increased with the incident neutron energy. Therefore, it is necessary to experimentally evaluate the transport of high-energy neutrons in shield materials. However, no experimental data was available for neutron transport in shield materials used in more than 100-GeV proton accelerator facilities that produce many high-energy neutrons.

We organized the Japanese and American Study of Muon Interaction and Neutron detection (JASMIN) with the objective of acquiring this experimental data. JASMIN began investigating the transport of high-energy neutrons within shields at an antiproton production (Pbar) target station of the Fermi National Accelerator Laboratory (FNAL), USA [2]. In the Pbar target station, the secondary particles produced by bombarding a thick target with 120-GeV protons are shielded by a thick iron and a thick concrete shield, as shown in Fig. 1. In our first experiment [3, 4], we installed Al, Cu, Nb, and Bi activation detectors on the outside of both the iron shield and the concrete shield. After accelerator operation, as indicators for neutron flux, the reaction rates of $^{27}$Al(n, α)$^{24}$Na, $^{209}$Bi(n, xn)$^{203-206}$Bi, $^{93}$Nb(n, 2n)$^{92m}$Nb, $^{nat}$Cu(n, x)$^{56-58}$Co, $^{nat}$Cu(n, x)$^{59}$Fe, and $^{nat}$Cu(n, x)$^{52, 54}$Mn in the activation detectors were nondestructively determined by γ-ray spectrometry. Because the nuclear reactions mentioned above mainly occur at neutron energies less than 100 MeV, we could indirectly measure the fluxes of less than 100-MeV neutrons passing through the activation detectors. The measured reaction rates were compared with those simulated using two calculation codes, PHITS [5] and MCNPX [6]. The results calculated using MCNPX exhibited a good agreement with the experimental results for the outsides of both the iron and the concrete shields. On the other hand, PHITS overestimated the reaction rates.

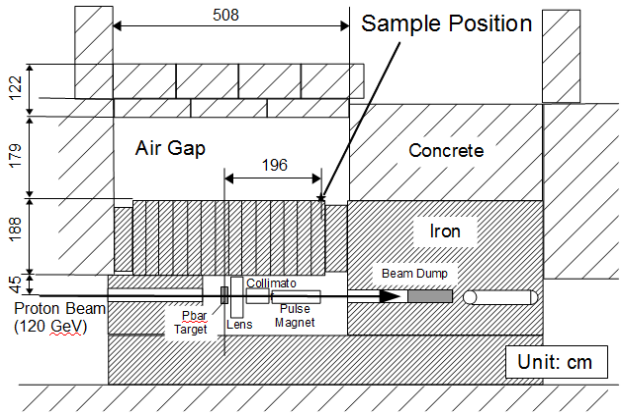

**Fig. 1.** Longitudinal-sectional views of Pbar target station taken along the proton beam. The star symbol indicates the experimental location of the Au activation detector.

At a large depth in a thick shield, low-energy neutrons having low penetrability are continually generated by the interactions of higher-energy neutrons with nuclei. Because high-energy neutrons having high penetrability play a vital role in the transport of neutrons within shields, it is important to measure neutrons having energies more than 100 MeV in order to clarify the problem in theoretical calculations. The Au activation method was developed to indirectly measure flux of more than 100-MeV neutrons in [7–9]. In this study, as indicators for flux of more than 100-MeV neutrons, production rates of spallation nuclides in Au activation detectors were measured close to one of the positions at which activation detectors were installed in a previous study [3] by using a low-background γ-ray counting system [10]. The measured results were compared with those calculated using PHITS.

## 2. EXPERIMENT

### 2.1 Au activation detector and neutron exposure

A stack of ten Au plates (dimensions: 20 × 20 × 0.5 mm$^3$; purity: 99.95%) was used as an Au activation detector. The stack was sealed in a polyethylene bag. The Au activation detector was installed on the outside of the iron shield at the Pbar target station in FNAL, as indicated by the closed star symbol in Fig. 1. The detector position was 196-cm downstream from the target axis. Therefore, the Au activation detector was located at a 50° angle from the beam axis. In the target room, the Pbar target (shape: cylindrical, diameter: 11.4 cm, height: 12.7 cm, material: inconel-600) was bombarded by 120-GeV protons. The practical thickness of the Pbar target along the beam axis was 8 cm. There existed a collection lens, a collimator, a magnet, and a beam dump downstream of the Pbar target. Particles from the Pbar target were shielded by a 188-cm-thick iron shield (ANSI M1020, density: 7.87 g/cm$^3$). An air gap exists behind the iron shield. The Au activation detector was exposed to neutrons for 20.16 days of accelerator operation. The total number of primary protons incident on the Pbar target was measured as 3.094 × 10$^{18}$ using a beam current monitor.

## 2.2 Determination of production rates

After neutron exposure, the radioactivities of $^{196}$Au and and $^{185}$Os in all Au plates were nondestructively determined by γ-ray spectrometry using a high-purity germanium (HPGe) detector in order to investigate the depth profiles of the production rates in own detector stack. Fig. 2 shows the depth profiles of the relative activities of $^{196}$Au and $^{185}$Os from the Pbar target side. The relative activities were normalized by the activities in the fifth plate. The radioactivities of $^{196}$Au and $^{185}$Os were constant regardless of the depth. It is indicated that the radioactivities were not affected by self-absorption or inter-increment of neutrons in the 5-mm-thick Au stack. Therefore, the production rates of the spallation nuclides in all plates are available for the evaluation of the high-energy neutron flux without correction for the self-absorption and inter-increment of neutrons.

Because the strong radioactivity of $^{198}$Au interferes with the γ-ray measurements, the plates on both sides of the stack were not used for this analysis. The second Au plate from the Pbar target side was used for nondestructive γ-ray spectrometry for $^{196}$Au, $^{188}$Pt, $^{189}$Ir, $^{185}$Os, and $^{175}$Hf. A low-background γ-ray counting system, with an HPGe detector coupled with three NaI(Tl) scintillation counters and four plastic scintillation counters [10], was used for γ-ray detection. Anticoincidence spectra with NaI(Tl) signals except for the X-ray region were used for the detection of the γ-rays because the coincidence of X-rays emitted immediately after electron capture reduced the peak-count recovery. The detector efficiencies were determined by using the γ-ray count rates from a γ-ray source, which was an identical Au plate activated by 370-MeV neutrons, at the Research Center for Nuclear Physics (RCNP) of Osaka University, and was determined radioactivities of $^{196}$Au, $^{188}$Pt, $^{189}$Ir, $^{185}$Os, and $^{175}$Hf using an HPGe detector previously calibrated by Canberra [11] and the LabSOCS software of Canberra [12].

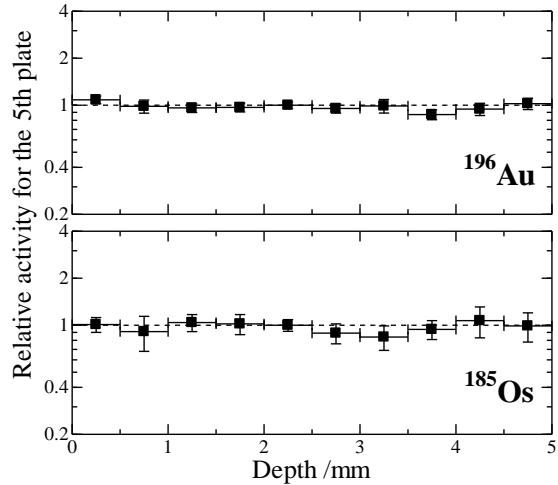

**Fig. 2.** Depth profiles of relative activities of $^{196}$Au and $^{185}$Os from the Pbar target side in the Au activation detector.

Because the radioactivities of $^{173}$Lu, $^{171}$Lu, and $^{169}$Yb were extremely low, five Au plates (plates 3–7 from the Pbar target side) were used together and the radiochemical determination of the radioactivities of $^{173}$Lu, $^{171}$Lu, and $^{169}$Yb was performed. Following dissolution of the set of Au plates in aqua regia, 20 mg of Nd$^{3+}$ carrier was added to the solution. Then, the rare earth elements were radiochemically separated as Nd$_2$(C$_2$O$_4$)$_3$. The low-background γ-ray counting system described above was also used for the γ-ray detection of $^{173}$Lu, $^{171}$Lu, and $^{169}$Yb from the Nd$_2$(C$_2$O$_4$)$_3$ sample. The γ-rays could be selectively detected by using the γ-X coincidence for $^{171}$Lu and the γ-X-X coincidence for $^{173}$Lu and $^{169}$Yb. The details of these γ-ray extractions and detector efficiencies are described in [10].

The production rates of the spallation nuclides were obtained from the measured radioactivity, number of detector nuclei, primary proton intensity, and duration of accelerator operation as the number of nuclei produced per Au atom and per primary proton. The intensity fluctuation of the primary protons was corrected by using the average intensity measured using the primary beam current monitor every 5 min.

## 3. RESULTS AND DISCUSSION

We could determine the production rates of 8 spallation nuclides ($^{196}$Au, $^{188}$Pt, $^{189}$Ir, $^{185}$Os, $^{175}$Hf, $^{173}$Lu, $^{171}$Lu, and $^{169}$Yb) in the Au activation detector. Fig. 3 shows a plot of the production rates measured in this study by closed squares as a function of the product mass number ($A_p$). The errors in the production rates refer to the counting statistics, detector efficiency (5%), and decay analysis. The solid line indicates an exponential trend that was obtained by the least-squares fits to the measured production rates.

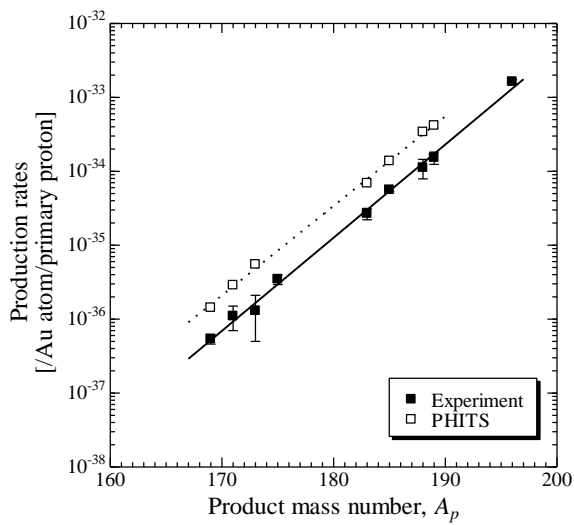

**Fig. 3.** Measured production rates (closed squares) and production rates calculated using PHITS (open squres) in the Au activation detector as a function of product mass number. The solid line and the dotted line indicate straight lines obtained by the least square fits to the measured and calculated production rates, respectively.

The measured production rates decreased exponentially with $A_p$. The slope of the decrease was quite steep. Because the production rate for smaller $A_p$ corresponds to the detection of higher-energy neutrons, the steep decrease observed at the Pbar target station indicates a small ratio of high-energy neutrons to low-energy neutrons. The iron shield efficiently reduced high-energy neutrons.

In order to compare the experimental and theoretical results, the neutron spectrum at 150–210 cm downstream from the Pbar target axis on the iron shield was calculated using the latest version of PHITS in the JAM mode. Here, a simple geometry, as described in [4], was used for the calculation. The production rate was calculated by the numerical integration at intervals of 1 MeV according to the following equation:

$$\text{Production rate} = \int_0^{120 \text{ GeV}} N(E_n) \times \sigma(E_p) dE,$$

where $E_n$ and $E_p$ are the energies of the neutron and proton, respectively, $N$ is the calculated number of neutrons per primary proton, and $\sigma$ is the cross section of a proton-induced reaction obtained from an excitation function based on abundant experimental database. The cross-section of the neutron reaction can be considered to be similar to that of proton reactions in the case of a complicated spallation reaction that emits more than nine nucleons. Therefore, we adopted the excitation functions of the proton reactions that produce the same nuclides as a substitute for those of the neutron reactions.

In Fig. 3, the production rates calculated using PHITS are plotted by open squares. The dotted line indicates an exponential trend that was obtained by the least-squares fits to the calculated production rates. The effective energies of $^{189}$Ir and $^{169}$Yb productions were estimated to be around 100–150 and 200–500 MeV. The calculated trend by the dotted line is similar to the experimental trend by the solid line. This result indicates that the profile of the calculated neutron spectrum agreed with the real neutron spectrum at more than 100 MeV. However, PHITS overestimated the production rates as same as in the low-energy reactions in previous study [3, 4]. Therefore, this result indicates that PHITS overestimated the flux at a neutron energy ranging from 100 MeV to around 500 MeV.

## 4. CONCLUSIONS

By using the Au activation method coupled with the low-background γ-ray counting system, we could evaluate the more than 100-MeV neutron spectrum calculated at the outside of the iron shield at an angle of 50° at the Pbar target station in FNAL. We proved that the Au activation method can serve as a powerful tool for indirect measurements of high-energy neutrons that play a vital role in neutron transport. The results measured in this study will be important for clarifying the problems in theoretical calculations of high-energy neutron transport.

## ACKNOWLEDGMENTS

This work was supported by a Grant-in-Aid from the Ministry of Education (KAKENHI 19360432 and 20354764), Japan. Fermilab is a U.S. Department of Energy Laboratory operated under Contract DE-AC02-07CH11359 by Fermi Research Alliance, LLC.